\documentclass[%
 reprint,
superscriptaddress,
 amsmath,amssymb,
 aps,
]{revtex4-2}
\usepackage{caption} % For caption spacing
\usepackage[labelformat=parens,justification=justified]{subcaption} % For sub-figures
\usepackage{ragged2e} % for the \justifying macro
\DeclareCaptionJustification{justified}{\justifying}
\usepackage{graphicx}
\usepackage{dcolumn}% Align table columns on decimal point
\usepackage{bm}% bold math
\usepackage{hyperref}% add hypertext capabilities
\begin{document}

\preprint{APS/123-QED}

\title{Two-dimensional Closed-Form Analytical Model of Laterally-Excited Film Bulk Acoustic Wave Resonator Multiferroic Antennas}% Force line breaks with \\

\author{Louis-Charles Ippet-Letembet}
\author{Rui-Fu Xu}
\author{Robin Jeanty}
\affiliation{Graduate Institute of Communication Engineering, National Taiwan University, Taipei 10617, Taiwan}
\author{Zhi (Jackie) Yao}
\affiliation{Applied Mathematics and Computational Research Division, Lawrence Berkeley National Laboratory, Berkeley, California 94720, USA}%
\author{Robert N. Candler}
\affiliation{California NanoSystems Institute, Los Angeles, California 90095, USA}
\affiliation{Department of Electrical and Computer Engineering, University of California, Los Angeles, California 90095, USA}
\author{Shih-Yuan Chen}
\altaffiliation{shihyuan@ntu.edu.tw}%Lines break automatically or can be forced with \\
\affiliation{Graduate Institute of Communication Engineering, National Taiwan University, Taipei 10617, Taiwan}
\affiliation{Department of Electrical Engineering, National Taiwan University, Taipei 10617, Taiwan}

\date{\today}% It is always \today, today,
             %  but any date may be explicitly specified

\begin{abstract}
To overcome the physical limitations of electrically small antennas, strain-mediated magnetoelectric antennas have been studied experimentally and theoretically. However, current closed-form analytical models include solely one-dimensional approaches. This paper proposes a two-dimensional closed-form analytical model of a laterally-excited multiferroic antenna. To model thickness-shear vibrations, we propose an extended Mason model to two dimensions. This theory solves unidirectionally-coupled elastodynamics-magnetization dynamics-electrodynamics. Without using the infinitesimal dipole approximation, this model provides a closed-form solution of the radiated electromagnetic field with an explicit dependency of the antenna response on the physical parameters, enabling their optimization. Furthermore,  we compare the efficiency-bandwidth product of different multiferroic antennas with Chu's limit in the Akhiezer regime, which delimits the highest achievable mechanical quality factor for acoustic resonators at GHz frequencies. We show that phonon-phonon coupling of Akhiezer damping is most likely to limit multiferroic antenna radiation, not Chu's limit. Moreover, judicious choice of materials can enhance the maximum efficiency by more than an order of magnitude.
\end{abstract}

%\keywords{Suggested keywords}%Use showkeys class option if keyword
                              %display desired
\maketitle

%\tableofcontents

\section{Introduction}

The increasing demand for short-range sensors for biomedical applications, portable wireless devices for monitoring applications, and long-range sensors for extremely low-frequency communication systems has led the path for antenna miniaturization. However, traditional electrically small resonant antennas suffer from physical limitations such as low efficiency and bandwidth, and the platform effect due to the proximity with the ground plane. These limitations have caused antennas to be the constraining factor on many systems over the past several decades \cite{wheeler,Chu,Mclean}. 

In recent years, mechanical antennas have been developed based on a mechanically induced radiation mechanism. In this mechanism, static electric charges, electric dipoles, or magnetic dipoles are accelerated through mechanical motion, potentially alleviating the difficulties associated with traditional electrically small antennas \cite{Hassanien,Chen,Schneider}. Technologies employ rotating permanent magnets \cite{Glickstein}, rotating electrets \cite{Cui}, or ferromagnetic-piezoelectric heterostructures \cite{Nan,Zaeimbashi,Dong}. 

Among these new kinds of antennas, the latter, also referred to as multiferroic antennas, when integrated into film bulk acoustic wave resonator (FBAR), have shown to be of particular interest for GHz frequencies applications \cite{Nan,Zaeimbashi,Ruifu}. The radiation source is the dynamic magnetization (spin wave) in the ferromagnetic thin film generated through the piezoelectrically induced magnetoelectric effect. Modeling such an antenna implies solving a system of coupled equations that include elastodynamics, magnetization dynamics, and electrodynamics.

Domann and Carman first proposed a one-dimensional closed-form model of strain-powered rod antennas, composed of a single bulk piezoelectric material or a single bulk piezomagnetic material \cite{Domann}. However, radiation-induced from ferromagnetic materials is not in the scope of such a model where magnetization dynamics is not considered.  Moreover, the model cannot be applied to multilayer structures and neglects the additional material required for the excitation of ferromagnetic materials. Later, an analytical study used the classical Mason model to study the radiation by a multilayer piezoelectric-piezomagnetic structure at Very-Low-Frequency \cite{Du}. The circuit model is one-dimensional and employed for obtaining the impedance characteristic of the multiferroic antenna. Nevertheless,  magnetization dynamics is not included, and the radiation is calculated by combining piezomagnetic and piezoelectric constitutive equations with the dipole approximation. Luong and Wang investigated the dynamic magnetoelastic coupling in multiferroic antennas, where for different classes of magnetic materials, the impact of the applied stress orientation was analyzed for optimal coupling \cite{Luong}. However, the study neglects elastodynamics,  and the magnetization is thus only time-dependent. Furthermore, electrodynamics is also not considered and no quantification of the radiation is proposed. 

Later, numerical methods have been proposed to study bulk acoustic wave (BAW) multiferroic antennas. Yao et al. introduced a three-dimensional unconditionally stable finite-difference time-domain (FDTD) algorithm for the study of BAW multiferroic antennas near ferromagnetic resonance (FMR) through the simultaneous solving of the coupled Landau-Lifshitz-Gilbert/Newton's/Maxwell's equations system \cite{Yao}. Ji et al. used a one-dimensional FDTD algorithm to investigate the impact of magnon-phonon coupling on the antenna polarization \cite{Ji}. However, because the physics phenomena considered intervene at very different size scales, numerical algorithms require a large computational time to solve the fully coupled partial differential equations system. A series of Finite-Element-Methods (FEM) based studies have later been proposed to analyze multiferroic antennas under thickness excitation of the piezoelectric \cite{Cai,Xu,Li,Gan}. There are methods using the commercially available software COMSOL where magnetostrictive and piezoelectric effects are analyzed separately with the magnetostrictive and the piezoelectric modules of the software. The radiated electromagnetic (EM) field is then calculated with the dipole approximation. Contrary to FDTD algorithms, these FEM-based studies using COMSOL do not solve the system of governing equations simultaneously and therefore neglect the mutual coupling between the governing equations. In doing so, they do not use the full potential of a 3D solution while still keeping its disadvantages, namely the very lengthy computational time. 

We propose here a two-dimensional closed-form analytical model of multiferroic antennas integrated into a laterally excited FBAR (XBAR). The XBAR generates thickness-shear modes, implying mechanical vibrations in both thickness and lateral directions. A Mason model extended to two dimensions is proposed to account for the mechanical wave propagation in both directions. Considering a unidirectional coupling between the governing equations, the knowledge of the mechanical displacement in the ferromagnetic layer enables us to obtain a closed-form analytic solution of the space-time-dependent spin wave. The spin wave is plugged into the vector potential as a source term and allows us to calculate the radiated EM field without infinitesimal dipole approximation. The closed-form solutions explicitly specify the parametric relationships, leading the path for an optimization of these parameters. Finally, the efficiency-bandwidth product of different multiferroic antennas is compared with Chu's limit in the Akhiezer regime. The Akhiezer regime refers to the phonon-phonon scattering, which sets for an acoustic resonator, the highest mechanical quality factor achievable in the GHz frequencies. We show that with a proper selection of piezoelectric and ferromagnetic materials, multiferroic antennas operating at the Akhiezer limit can approach Chu's limit.

\section{Theoretical description of laterally excited multiferroic antennas}

\subsection{Operating principle of multiferroic antennas}

The multiferroic antenna consists of a thin-film piezoelectric material deposited on a silicon substrate with a pair of electrodes placed along the x-axis. On top of the piezoelectric is deposited a thin-film ferromagnetic material. The silicon substrate is etched away to create an air cavity beneath the piezoelectric, enabling vibrations in both lateral and thickness directions. As shown in Fig. 1(a), the active region studied in this analytical model is the free-standing piezoelectric-ferromagnet structure. The active region is delimited by the length $(L)$ of the ferromagnet and the thickness of the antenna. The boundary conditions along the thickness and lateral directions are stress-free conditions. For comparison and verification, the structure is also analyzed using COMSOL, as shown in Fig. 1(a), where both electrodes and piezoelectric material combined with perfectly matched layers (PML) are considered. Aluminum Nitride (AlN), is used as piezoelectric material for its high longitudinal acoustic velocity. Nickel (Ni) is used as ferromagnetic material and Platinum (Pt) for the electrodes. When a time-varying voltage is applied to the electrodes, the piezoelectric is excited and generates acoustic waves in the form of thickness-shear waves. The acoustic waves propagate into the ferromagnetic layer, generating time-harmonic standing spin waves, the source of EM radiation. 

The dynamic response of the multiferroic antenna is predicted with a set of equations comprising  Maxwell's equations for EM radiation, the Landau-Lifshitz (LL) equation for magnetization dynamics, and Newton's Law for acoustic waves:
\begin{equation}
    \begin{array}{ll}
        \nabla\times \mathbf{E}=-\frac{\partial \mathbf{B}}{\partial t} &,~\nabla\times \mathbf{H}= \mathbf{J}+ \frac{\partial \mathbf{D}}{\partial t}
        \end{array}
\end{equation}
\begin{equation}
    \frac{\partial \mathbf{m}}{\partial t}=- \mu_0\gamma\left(\mathbf{m}\times \mathbf{H_{eff}}\right)-\alpha_L\mu_0\gamma \mathbf{m}\times(\mathbf{m}\times \mathbf{H_{eff}})
\end{equation}
\begin{equation}
    \begin{array}{ll}
       \rho\frac{\partial^2 \mathbf{u}}{\partial t^2}=\nabla \cdot \boldsymbol{\sigma}  &,~ \boldsymbol{\varepsilon}=\frac{1}{2}\left(\left(\nabla \mathbf{u}\right)^T+\nabla \mathbf{u}\right)
    \end{array}
\end{equation}

In (1), $\mathbf{E},\mathbf{H}$ are the electric field and the magnetic field, $\mathbf{D},\mathbf{B}$ are the electric displacement field and the magnetic flux density, and $\mathbf{J}$ is the free current. In (2), $\mathbf{m}$ is the magnetization unit vector defined as $\mathbf{m}=\frac{\mathbf{M}}{M_s}$, with $\mathbf{M}$ the magnetization vector, $M_s$ the saturation magnetization. $\mathbf{H_{eff}}$ is the effective magnetic field composed of the different physics phenomena that force the magnetization vector to rotate. $\mu_0$ is the vacuum permeability, $\gamma$ is the gyromagnetic ratio with a value of $-1.76\times10^{11}C\cdot kg^{-1}$, and $\alpha_L$ is a dimensionless damping constant with a value of $0.025$ \cite{Bhagat}. In (3), $\boldsymbol{\sigma},\boldsymbol{\varepsilon}$ are the stress and strain fields. $\mathbf{u}$ is the displacement vector and $\rho$ the density. The superscript $T$ refers to the transpose.

Hooke's law relates the stress and strain fields through the material elastic stiffness tensor $C$ and the viscosity tensor $\eta$. For a non-piezoelectric material:
\begin{equation}
\boldsymbol{\sigma}=C:\boldsymbol{\varepsilon}+\eta\frac{\partial\boldsymbol{\varepsilon}}{\partial t}
\end{equation}

In the piezoelectric, Hooke's law takes the form of the so-called piezoelectric constitutive equations:
\begin{equation}
    \begin{split}
        &\boldsymbol{\sigma}=C:\boldsymbol{\varepsilon}+\eta\frac{\partial\boldsymbol{\varepsilon}}{\partial t}-e^T\cdot \mathbf{E} \\
        &\mathbf{D}=e:\boldsymbol{\varepsilon}+\epsilon \mathbf{E}
    \end{split}
\end{equation}
where $e$ is the piezoelectric constant in the stress-charge formulation, and $\epsilon$ is the piezoelectric permittivity.

The system of equations is solved in a unidirectional way. First, we solve Newton's law combined with the extended two-dimensional Mason model. Only the Villari effect (inverse magnetostriction) is considered in this work. Thus, the displacement vectors obtained through elastodynamics enable us to find the expression of the magnetoelastic field, the major source of magnetization rotations. Subsequently, the LL equation is solved which gives the spin wave. Finally, the spin wave is plugged into Maxwell's equations as a source term to obtain the EM far-field.

\subsection{Elastic waves}
The electrodes placed along the x-axis generate particle displacement on the xz-plane, namely thickness-shear modes. The dominant stress in the structure is the shear stress in the same plane. The shear stress vanishes at the lateral boundaries at $x=0$ and $x=L$, which leads to the following general form for the x and z components of the displacement vector in the structure:
\begin{equation}
    \begin{split}
        &u_x=\sum_{n=1}^\infty u_{xn}\left(u_{x}^{+}e^{-jk_zz}+u_{x}^{-}e^{jk_zz}\right)\sin\left(k_nx\right)\\
        &u_z=\sum_{n=1}^\infty u_{zn}\left(u_{z}^{+}e^{-jk_zz}+u_{z}^{-}e^{jk_zz}\right)\cos\left(k_nx\right)
    \end{split}
\end{equation}

The propagation constant along the lateral direction x is directly fixed in each layer by the boundary conditions and its value is $k_n=\frac{(2n-1)\pi}{L}$ where $n\in \mathbb{N}$. $u_{xn}$ and $u_{zn}$ are the eigenvectors. The piezoelectric and the ferromagnetic materials have different mechanical properties, rendering the bilayer structure inhomogeneous. As a consequence, the wavenumber $k_z$ along the thickness direction varies from one layer to another. It will be denoted as $k_p$ in the piezoelectric and $k_f$ in the ferromagnetic layer. The two propagation constants along the thickness direction cannot be determined merely by the application of boundary conditions. They are determined by the dispersion relations. Inserting $u_x$ and $u_z$ in Newton's Law and examining individual modes gives the dispersion relations in each layer (See Supplemental Material \cite{Supp}).

\begin{widetext}
    \begin{eqnarray}
   &&\omega^2=\frac{\left(\underline{C_{11}^f}+\underline{C_{44}^f}\right)\left(k_n^2+k_f^2\right)}{2\rho}+\frac{\sqrt{\left(\underline{C_{11}^f}-\underline{C_{44}^f}\right)^2\left(k_f^2-k_n^2\right)^2+4k_n^2k_f^2\left(\underline{C_{44}^f}+\underline{C_{12}^f}\right)^2}}{2\rho}\\
   &&
    \omega^2=\frac{\left(\underline{C_{11}^p}+\underline{C_{44}^p}\right)k_n^2+\left(\underline{C_{33}^p}+\underline{C_{44}^p}\right)k_p^2}{2\rho}\left(1+\sqrt{1-\frac{\left(\underline{C_{11}^p}k_n^2+\underline{C_{44}^p}k_p^2\right)\left(\underline{C_{33}^p}k_p^2+\underline{C_{44}^p}k_n^2\right)-\left(\underline{C_{13}^p}+\underline{C_{44}^p}\right)^2k_n^2k_p^2}{\left(\underline{C_{11}^p}+\underline{C_{44}^p}\right)k_n^2+\left(\underline{C_{33}^p}+\underline{C_{44}^p}\right)k_p^2}}\right)
\end{eqnarray}
\end{widetext}

In (7) and (8), $\omega$ is the frequency of the AC voltage applied at the electrodes. $\underline{C_{ij}^f},\underline{C_{ij}^p}$ are the elastic stiffness constants in the ferromagnetic and the piezoelectric layers, respectively, when mechanical damping is considered:

\begin{equation}
    \underline{C_{ij}^{f,p}}=C_{ij}^{f,p}\left(1+j\eta_s\right)
\end{equation}
where $\eta_s$ is the viscosity constant. The dispersion relations relate known quantities, the frequency of the AC applied voltage $\omega$ and the wavenumber $k_n$ in the lateral direction, to the unknown wavenumbers along the thickness direction $(k_p,k_f)$. Therefore, the latter can be extracted from the dispersion relations. However, the two dispersion relations taken separately cannot uniquely determine the structure's resonance frequencies. A tool is thus needed to obtain the resonance frequencies of such a multi-layer structure, and this is accomplished by the proposed extended Mason model. In the study of BAW resonators, conventionally, the Mason model has only been employed as a 1D approximation for obtaining the structure's electrical response: the resonance frequencies through the input impedance or the reflectivity of acoustic mirrors for instance \cite{Hashimoto}. The 1D analysis is then combined with COMSOL for the study of mode shapes \cite{Thalhammer}. On the contrary, we show here that extended to two dimensions and combined with Newton's law, an extended Mason model can give both electrical response and mode shapes.

\subsection{Extended Mason model: electrical response}
We define an extended Mason model per unit length and extend it to two dimensions taking advantage of the coupling in Newton's law between the x and z components of the displacement vector. Suppose we project Newton's equation onto the z-axis and examine the waves propagating along the positive and negative thickness directions. In that case, one can find the coupling relation between the magnitudes along x and z. In the piezolayer for instance (See Supplemental Material \cite{Supp}):
\begin{equation}
    \left[\begin{array}{l}
         u_{z}^+  \\
         u_{z}^- 
    \end{array}\right]=jU_p\left[\begin{array}{c}
         -u_{x}^+  \\
         u_{x}^- 
    \end{array}\right]
\end{equation}
where $U_p$ is the coupling constant defined by:
\begin{equation}
   U_p=\frac{k_pk_n\left(\underline{C_{44}^p}+\underline{C_{13}^p}\right)}{-\rho\omega^2+\underline{C_{33}^p}k_p^2+\underline{C_{44}^p}k_n^2}
\end{equation}

In other words, the coupling constant $U_p$ allows a circuit representation of a 2D structure where the wave propagation in one direction is considered first and the wave propagation in the other direction is directly deduced by (10). Fig. 1(b) is the extended Mason model for the present bilayer resonator under lateral field excitation. The left part of the circuit with an electrical port represents the piezolayer, and the right part of the circuit is the ferromagnetic layer. The stress-free boundary conditions at the top and bottom of the active region are represented in the circuit by two acoustic shorts. One at the bottom of the piezolayer, denoted by the plane $z=0$, and the other at the top of the ferromagnetic layer denoted by the plane $z=z_2$. The plane $z=z_1$ is the piezoelectric-ferromagnet interface. In the circuit, the voltages are the modal shear forces denoted as $F(z)$.

\begin{figure*}
\begin{subfigure}{0.4\linewidth}
\caption{}
  \label{fig:sfig1}
  \includegraphics[width={\linewidth}]{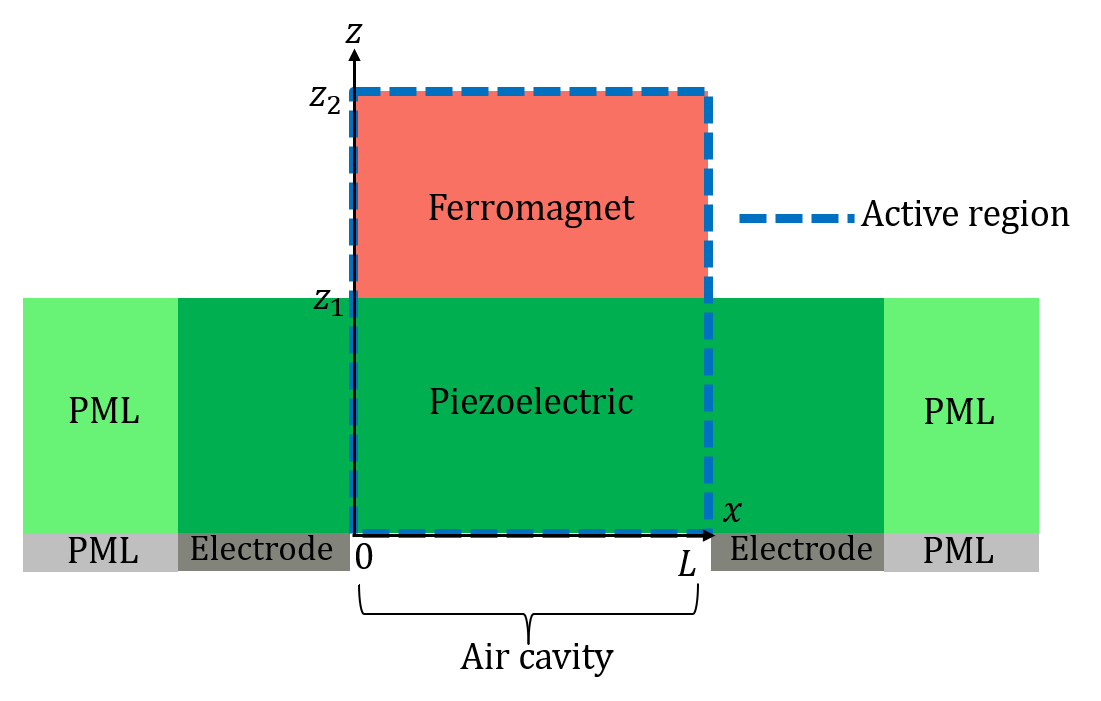}  
\end{subfigure}\hspace{5pt}
\begin{subfigure}{0.5\linewidth}
  \caption{}
  \label{fig:sfig2}
  \includegraphics[width={\linewidth}]{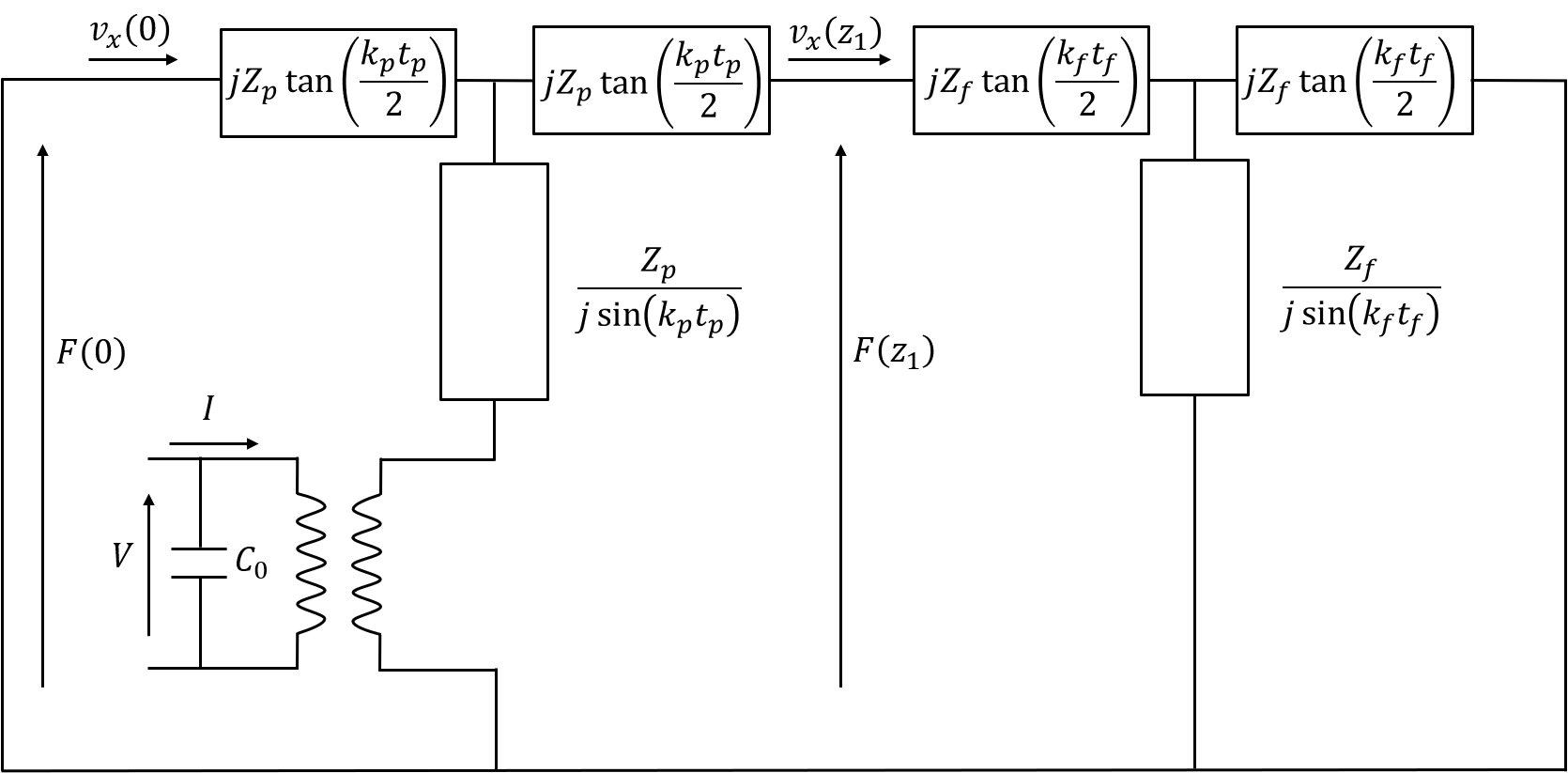}
\end{subfigure}
\begin{subfigure}{0.45\linewidth}
  \centering
  \caption{}
  \label{fig:sfig3}
  \includegraphics[width={\linewidth}]{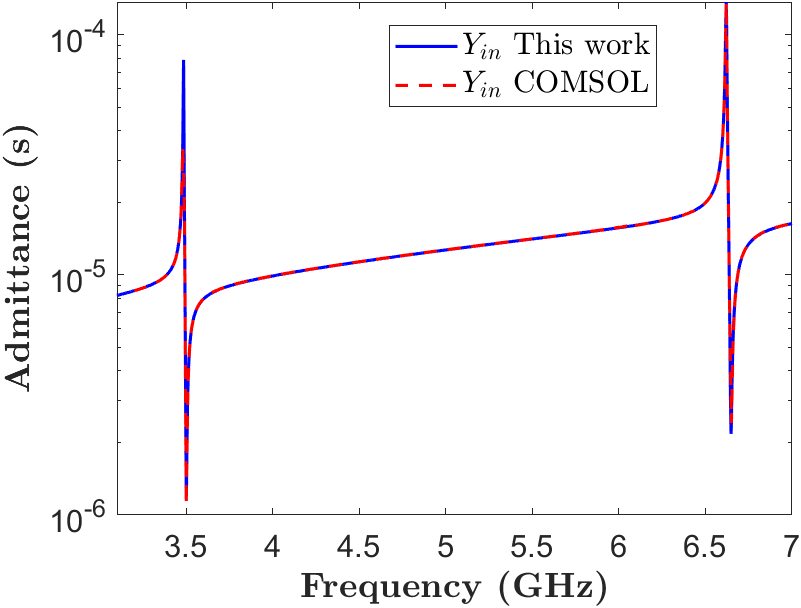}
\end{subfigure}\hspace{8pt}
\begin{subfigure}{0.45\linewidth}
  \centering
  \caption{}
  \label{fig:sfig4}
  \includegraphics[width={\linewidth}]{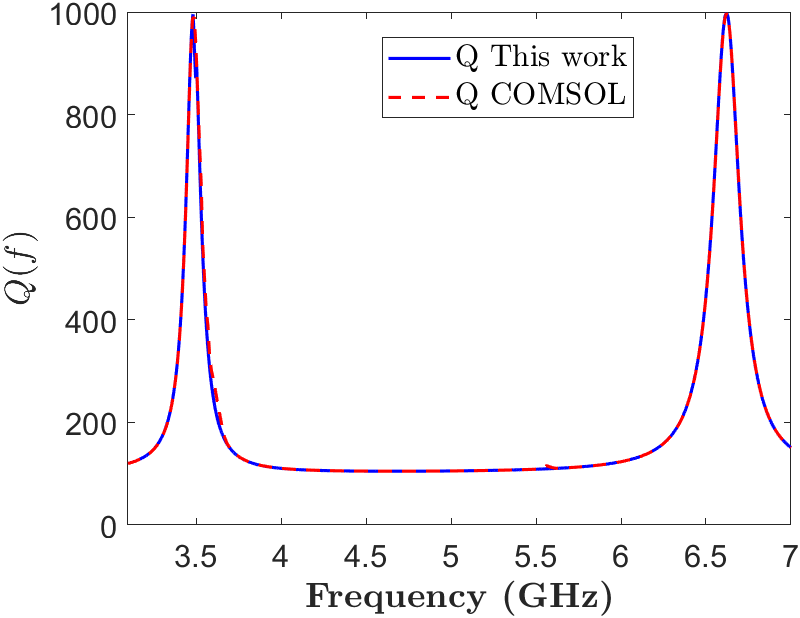}
\end{subfigure}
\caption{(a) 2D Schematic representation of the FBAR antenna.  (b) Two-dimensional extended Mason model. Comparison of the input admittance and quality factor obtained with COMSOL and the model in (c) and (d).}
\label{fig:fig}
\end{figure*}

Owing to the orthogonality of the mechanical waves, we modify the definition of the shear force by taking the weighted integral along the x-axis of the shear stress. In the piezolayer:
\begin{equation}
\begin{split}
    F(z)&=-t_p\int_0^{L}\sin\left(k_nx\right)\sigma_{xz}dx\\
        &=Z_p\left(j\omega u_{x}^+e^{-jk_pz}-j\omega u_{x}^-e^{jk_pz}\right)+\frac{2e_{15}Vt_p}{(2n-1)\pi}
\end{split}
\end{equation}
where $t_p=z_1$ is the thickness of the piezolayer, $V$ the time-varying voltage applied to the electrodes, and $e_{15}$ the piezoelectric coupling constant for an electric field applied along the x-axis. Defining the modal shear force as a weighted integral allows two things. First to quantify how much each mode couples with the applied electric field in the lateral direction, represented by the coefficient $\frac{2}{(2n-1)\pi}$. Second to account for the coupling between the waves propagating in the thickness and lateral directions in the definition of the mechanical impedance $Z_p$:  
\begin{equation}
    Z_p=\frac{Lt_p}{2}\frac{\underline{C_{44}^p}k_p}{\omega}\left(1-\frac{k_nU_p}{k_p}\right)
\end{equation}

In the circuit, the currents are the magnitude of the acoustic velocities along the lateral direction denoted as $v_{x}(z)$.
Applying Kirchhoff's laws, we can solve for the magnitudes of the acoustic velocities at the planes $z=0$ and $z=z_1$:  
\begin{equation}
    \begin{split}
        v_x(0)&=-jv\left(Z_f\tan\left(k_ft_f\right)+Z_p\tan\left(\frac{k_pt_p}{2}\right)\right)\\
        v_x(z_1)&=jvZ_p\tan\left(\frac{k_pt_p}{2}\right)\\
        v&=\frac{2e_{15}t_p}{(2n-1)\pi}\frac{V}{Z_p^2\left(\frac{Z_f\tan\left(k_ft_f\right)}{Z_p\tan\left(k_pt_p\right)}+1\right)}
    \end{split}
\end{equation}
where $t_f=z_2-z_1$ and $Z_f$ are the ferromagnetic layer's thickness and mechanical impedance, respectively. (14) gives the magnitude of the acoustic velocities of the n-th propagating mode at the top and bottom of the piezolayer with respect to the applied voltage $V$. The modal current is the surface integral of the modal displacement field along the thickness and width of the piezolayer \cite{Rosenbaum}:
\begin{equation}
    I_n=j\omega\int_0^{W_p}\int_{0}^{z_1}Ddydz
\end{equation}
where $W_p$ is the width of the piezolayer. Note that the modal current is a $\sin\left(k_nx\right)$ function. Therefore, we modify the classic Mason model again by taking the modal current's mean value. It enables to take into account the contribution of the propagation in the lateral direction to the modal current's magnitude:
\begin{equation}
    \Tilde{I}_n=\frac{1}{L}\int_0^L I_ndx 
\end{equation}

The input admittance $Y_{in}$ can be obtained by expressing the total current $I$ as a function of the applied voltage $V$. The total current $I$ is defined as the sum of the static current, related to the piezoelectric static capacitance, and the current flowing in the motional branches. The latter is the sum over all the modes of the mean current $\Tilde{I}_n$. The input admittance $Y_{in}$ can now be written as:
\begin{equation}
        Y_{in}=j\omega C_0\left(1+\sum_{n=1}^\infty k_{eff,n}^2\right)
\end{equation}
where $C_0=\frac{W_pt_p\epsilon}{L}$ is the piezoelectric static capacitance, and $k_{eff,n}^2$ is the effective electromechanical coupling coefficient of the n-th mode that can be expressed as:
\begin{equation}
    k_{eff,n}^2=k_x^2\frac{K^2}{k_pt_p/2}\frac{\tan\left(\frac{k_pt_p}{2}\right)+\frac{Z_f}{2Z_p}\tan\left(k_ft_f\right)}{1+\frac{Z_f\tan\left(k_ft_f\right)}{Z_p\tan\left(k_pt_p\right)}}
\end{equation}
where $K^2=\frac{e_{15}^2}{\epsilon C_{44}^p}$. Note here that $k_x^2=\frac{8}{\pi^2}\frac{1}{(2n-1)^2}$ corresponds to the lateral component and follows from the Berlincourt's formula \cite{Berlincourt}:
\begin{equation}
    \begin{split}
        k_x^2&=\frac{\left(\int_0^L E_x\sin{\left(k_nx\right)\mathrm{d}x}\right)^2}{\left(\int_0^LE_x^2\mathrm{d}x\right)\left(\int_0^L\sin{\left(k_nx\right)}^2\mathrm{d}x\right)}\\
        &=\frac{8}{\pi^2}\frac{1}{(2n-1)^2}
    \end{split}
\end{equation}
where $E_x$ is the electric field induced by the applied voltage along the x-axis. The input admittance is now defined. The mechanical quality factor $Q$ of the XBAR multiferroic antenna can be expressed as \cite{Jin}:
\begin{equation}
        Q=\omega\bigg\lvert\frac{dS_{11}}{d\omega}\bigg\rvert\frac{1}{1-\lvert S_{11}\rvert^2}
\end{equation}
where $S_{11}$ can be expressed in terms of the input admittance as:
\begin{equation}
    S_{11}=\frac{1-Y_{in}Z_s}{1+Y_{in}Z_s}
\end{equation}
where $Z_s=50$ $\mathrm{\Omega}$ is the system impedance.

Material properties of AlN and Ni layers, along with the Pt electrodes, are tabulated in Table I. The thickness of the AlN and Ni layers are 400 and 300 nm respectively. The width of the AlN layer is $W_p=5$ mm and the length of the active region is  $L=400$ $\mu m$. In COMSOL, the electrodes' thickness and length are 30 nm and 200 $\mu m$ respectively. The  Anchor losses are modeled in COMSOL and this model with a common viscosity constant for the piezoelectric and ferromagnetic materials of $\eta_s=0.001$. The dielectric losses are also taken into account with a loss tangent of $\tan\delta=0.01$. 

Fig. 1(c) shows the frequency responses of the input admittance obtained with this model and with COMSOL simulation.  The first two acoustic resonances are observed at 3.49 GHz and 6.63 GHz. Fig. 1(d) compares the quality factor of the XBAR multiferroic antenna obtained with this model and with COMSOL. At each resonance, the quality factor reaches a value of approximately 1000, corresponding to the inverse of the viscosity constant.

\subsection{Extended Mason model: mode shape}
We show here that the extended two-dimensional Mason model allows us to describe particle displacement in both the thickness and lateral directions. In the piezolayer, the relation between the displacement magnitude $(u_x^+,u_x^-)$ and the acoustic velocities magnitude $\left(v_{x}(0),v_{x}(z_1)\right)$ is (See Supplemental Material \cite{Supp}):
\begin{equation}
    j\omega\left[\begin{array}{l}
         u_{x}^+  \\
         u_{x}^- 
    \end{array}\right]=\frac{1}{2j\sin\left(k_pt_p\right)}\left[\begin{array}{l}
        e^{jk_pz_1}v_{x}(0)-v_{x}(z_1) \\
        -e^{-jk_pz_1}v_{x}(0)+v_{x}(z_1) 
    \end{array}\right]
\end{equation}

Recall that (14) defines the acoustic velocities magnitude and that the coupling condition (10) relates the magnitudes $(u_{z}^{+},u_{z}^{-})$ in the other direction. Therefore, the particle displacement along the thickness and lateral directions in the piezolayer can be expressed as:
\begin{equation}
    \begin{split}
        &\begin{split}
            u_x&=\sum_{n=1}^\infty \frac{ju_{xn}}{(2n-1)\pi}\left(a_1\sin(k_p(z-z_1))+a_2\sin(k_p z)\right)\\
    &\times\sin(k_nx)e^{j\omega t}
        \end{split}\\
        &\begin{split}
        u_z&=\sum_{n=1}^\infty \frac{u_{zn}U_p}{(2n-1)\pi}\left(a_1\cos(k_p(z-z_1))+a_2\cos(k_p z)\right)\\
        &\times\cos(k_nx)e^{j\omega t}
    \end{split}
    \end{split}
\end{equation}
where $a_1,a_2$ are constants  defined as:
\begin{equation}
    \begin{split}
        a_1&=\frac{e_{15}E_x\left( \frac{Z_f}{Z_p}\tan\left(k_ft_f\right)+\tan\left(\frac{k_pt_p}{2}\right)\right)}{\underline{C_{44}^p}\left(k_p-k_nU_p\right)\sin\left(k_pt_p\right)\left(\frac{Z_f\tan\left(k_ft_f\right)}{Z_p\tan\left(k_pt_p\right)}+1\right)}\\
        a_2&=\frac{e_{15}E_x\tan\left(\frac{k_pt_p}{2}\right)}{\underline{C_{44}^p}\left(k_p-k_nU_p\right)\sin\left(k_pt_p\right)\left(\frac{Z_f\tan\left(k_ft_f\right)}{Z_p\tan\left(k_pt_p\right)}+1\right)}
    \end{split}
\end{equation}

\begin{figure*}
\begin{subfigure}{0.32\linewidth}
\centering
\caption{}
  \label{fig:sfig1}
  \includegraphics[width={\linewidth}]{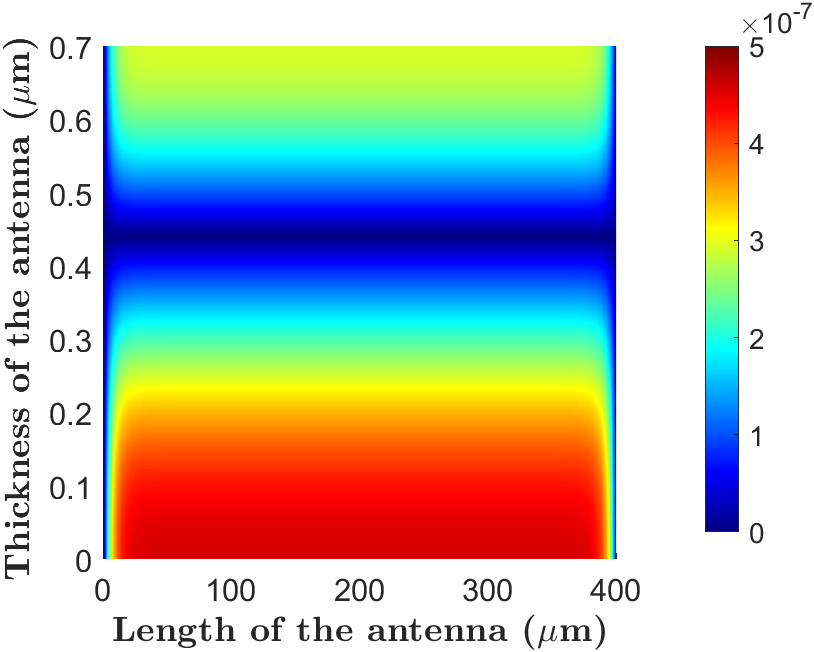}  
\end{subfigure}
\begin{subfigure}{0.32\linewidth}
  \centering
  \caption{}
  \label{fig:sfig2}
  \includegraphics[width={\linewidth}]{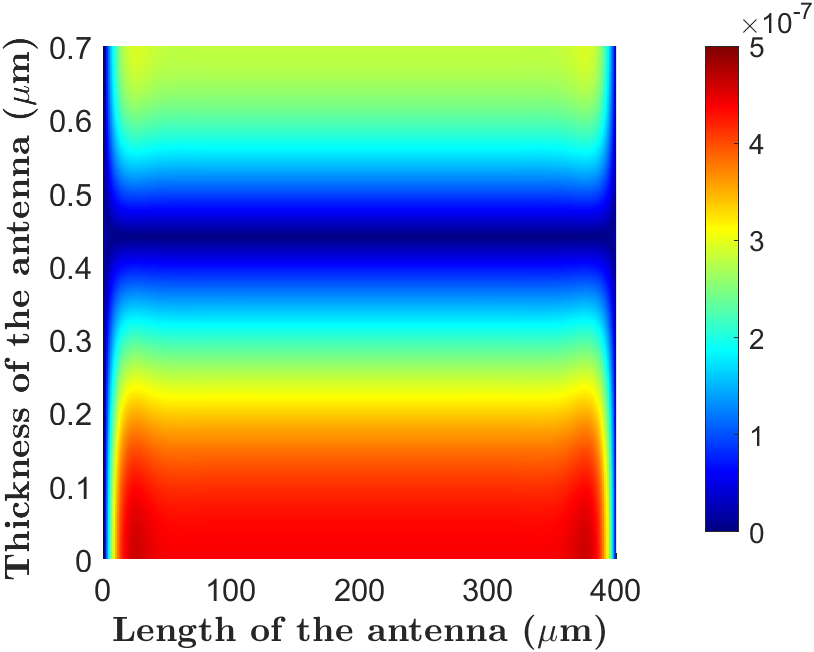}
\end{subfigure}
\begin{subfigure}{0.32\linewidth}
  \centering
  \caption{}
  \label{fig:sfig3}
  \includegraphics[width={\linewidth}]{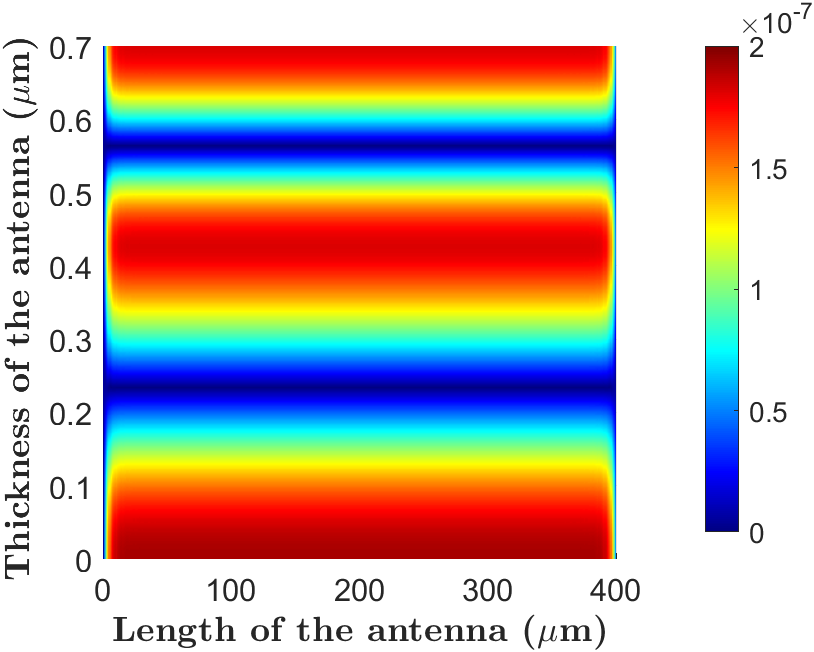}
\end{subfigure}
\begin{subfigure}{0.32\linewidth}
  \centering
  \caption{}
  \label{fig:sfig4}
  \includegraphics[width={\linewidth}]{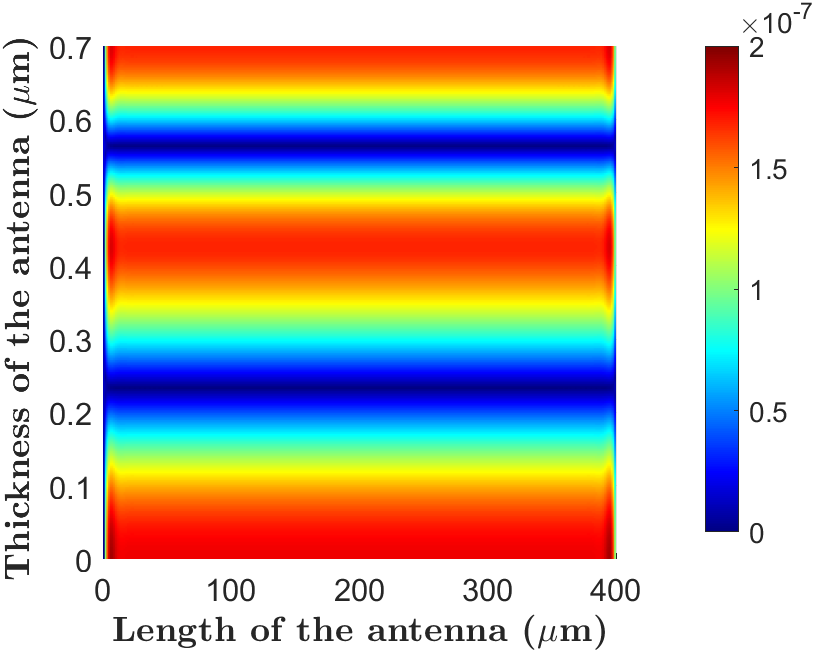}
\end{subfigure}
\begin{subfigure}{0.32\linewidth}
  \centering
  \caption{}
  \label{fig:sfig3}
  \includegraphics[width={\linewidth}]{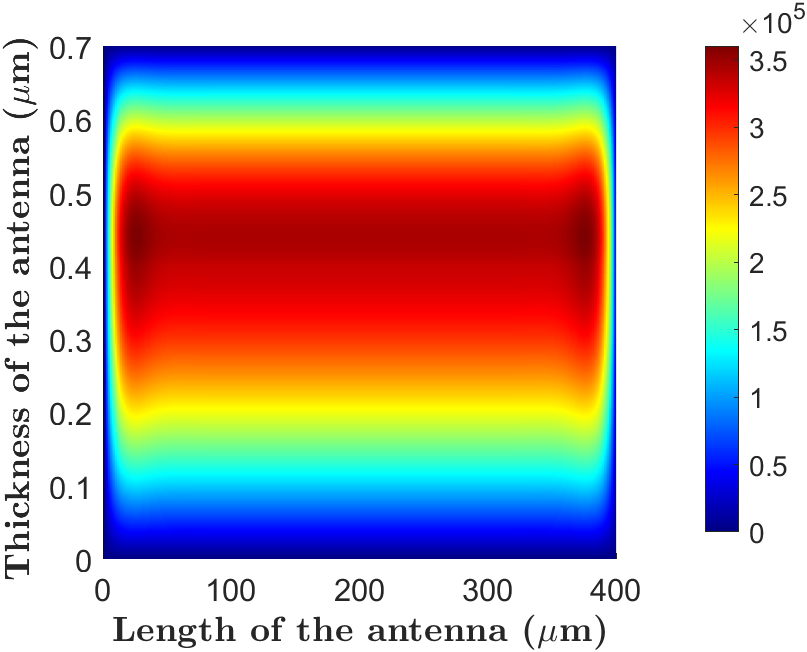}
\end{subfigure}
\begin{subfigure}{0.32\linewidth}
  \centering
  \caption{}
  \label{fig:sfig4}
  \includegraphics[width={\linewidth}]{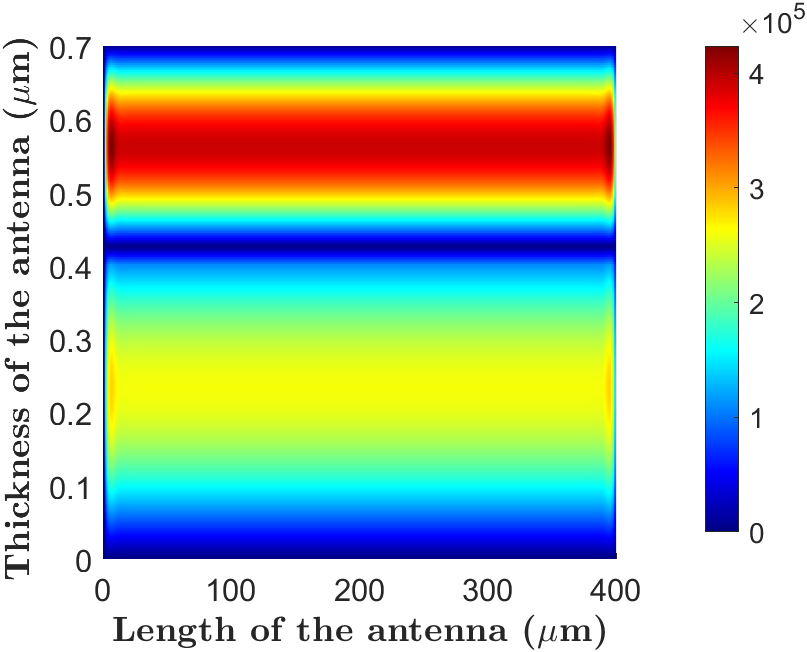}
\end{subfigure}
\caption{Displacement magnitude in $\mu m$ at first resonance with COMSOL in (a) and with this model in (b), and at second resonance with COMSOL in (c) and with this model in (d). Shear stress magnitude with this model in Pa at the first resonance in (e), and the second resonance in (f).}
\label{fig:fig}
\end{figure*}

The continuity of the stress is automatically considered in the circuit when Kirchhoff's laws are applied. As a consequence, only the continuity of the displacement vector at the piezoelectric-ferromagnet interface needs to be verified. Taking into account the stress-free condition at the top surface of the ferromagnetic layer leads to the following displacement vector components in the ferromagnetic layer:
\begin{equation}
    \begin{split}
        &u_x=\sum_{n=1}^\infty\frac{ju_{xn}}{(2n-1)\pi}a_3\cos\left(k_f(z-z_2)\right)\sin\left(k_n x\right)e^{j\omega t}\\
        &u_z=-\sum_{n=1}^\infty \frac{u_{zn}U_p}{(2n-1)\pi}a_4\sin\left(k_f(z-z_2)\right)\cos\left(k_nx\right)e^{j\omega t}
    \end{split}
\end{equation}
where $a_3, a_4$ are constants defined as:
\begin{equation}
    \begin{split}
        &a_3=\frac{e_{15}E_x\tan\left(\frac{k_pt_p}{2}\right)}{\underline{C_{44}^p}\left(k_p-k_n U_p\right)\cos\left(k_ft_f\right)\left(\frac{Z_f\tan\left(k_ft_f\right)}{Z_p\tan\left(k_pt_p\right)}+1\right)}\\
        &\begin{split}
            a_4&=\frac{e_{15}E_x}{\underline{C_{44}^p}(k_p-k_nU_p)\sin(k_pt_p)}\\
            &\times\frac{\frac{Z_f}{Z_p}\tan(k_ft_f)+\tan\left(\frac{k_pt_p}{2}\right)\left(1+\cos(k_pt_p)\right)}{\sin(k_ft_f)\left(\frac{Z_f\tan\left(k_ft_f\right)}{Z_p\tan\left(k_pt_p\right)}+1\right)}
        \end{split}
    \end{split}
\end{equation}

In each layer, the shear stress and strain in the xz-plane are derived using Hooke's law:
\begin{equation}
    \begin{split}
    \varepsilon_{xz}&=\frac{\partial u_x}{\partial z}+\frac{\partial u_z}{\partial x}\\
    \sigma_{xz}&=\sqrt{3}\underline{C_{44}^{f,p}}\varepsilon_{xz}
    \end{split}
\end{equation}
where we considered pure shear stress. In the present study, we considered the active region as a bilayer structure for simplicity. In the case where a structure with more than two layers is considered, the solution procedure of the problem is similar. Adding extra layers of non-magnetic buffer materials will shift the resonance frequencies and similarly to the present case, the continuity of the shear force would be automatically considered in the circuit. The continuity condition on the displacement vector solely needs to be applied to define in each layer the mode shape. 

Figs. 2(a)-2(d) plot in the active region, at the first two resonances, the magnitude of the displacement vector obtained with this model and with COMSOL. At the first resonance or 3.49 GHz, we observe zero displacements at $z\approx 450~\mathrm{nm}$. The displacement magnitude starts to increase when leaving the plane $z\approx 450~\mathrm{nm}$. It indicates that the shear force, proportional to the displacement gradient, is higher at this plane. At the second resonance or 6.63 GHz, the displacement gradient is higher in the middle of both layers, indicating that the shear force is maximized in the middle of each layer. This is indeed verified by the plots of the stress magnitudes at the first and second resonances shown in Figs. 2(e)-2(f). The maximum shear stress in the ferromagnetic layer reaches a value of 0.35 MPa for the first resonance and 0.42 MPa for the second resonance. As a consequence, the second mode will generate higher magnetization rotation in the ferromagnet through the magnetoelastic field, leading to higher radiation. 

The knowledge of mechanical displacement enables us to define the magnetoelastic field, the principal component of the effective magnetic field that generates the torque that rotates the magnetization. As will be detailed next, the spin wave is calculated analytically for both time and space dependencies with a unidirectional coupling considered. 

\begin{figure*}
\begin{subfigure}{0.41\linewidth}
  \centering
  \caption{}
  \label{fig:sfig1}
  \includegraphics[width={\linewidth}]{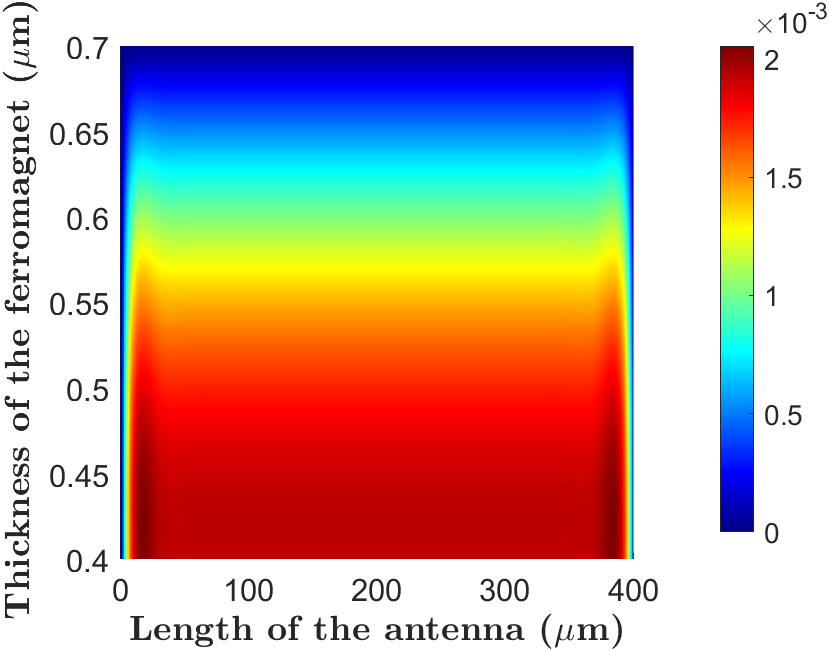}
\end{subfigure}\hspace{5pt}
\begin{subfigure}{0.41\linewidth}
  \centering
  \caption{}
  \label{fig:sfig2}
  \includegraphics[width={\linewidth}]{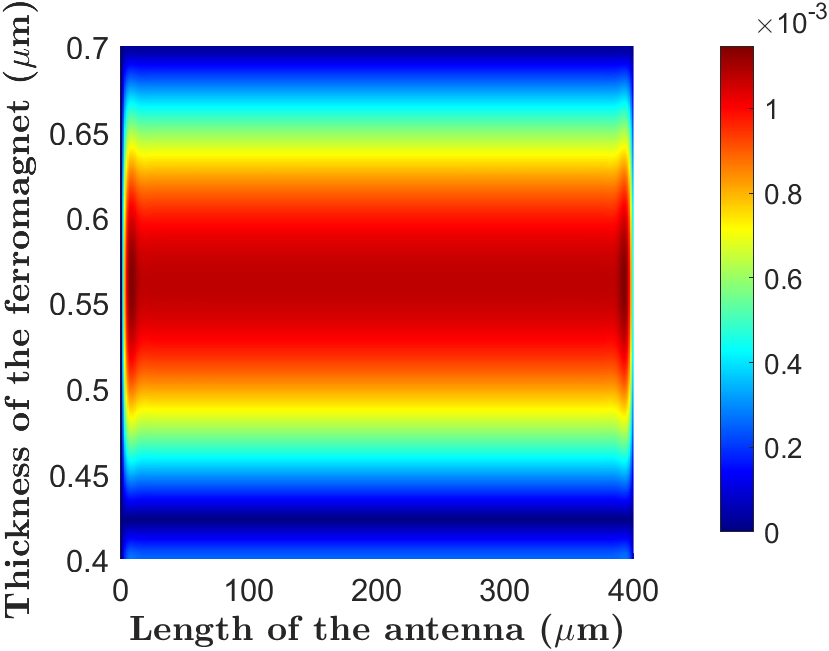}
\end{subfigure}
\begin{subfigure}{0.41\linewidth}
  \centering
  \caption{}
  \label{fig:sfig3}
  \includegraphics[width={\linewidth}]{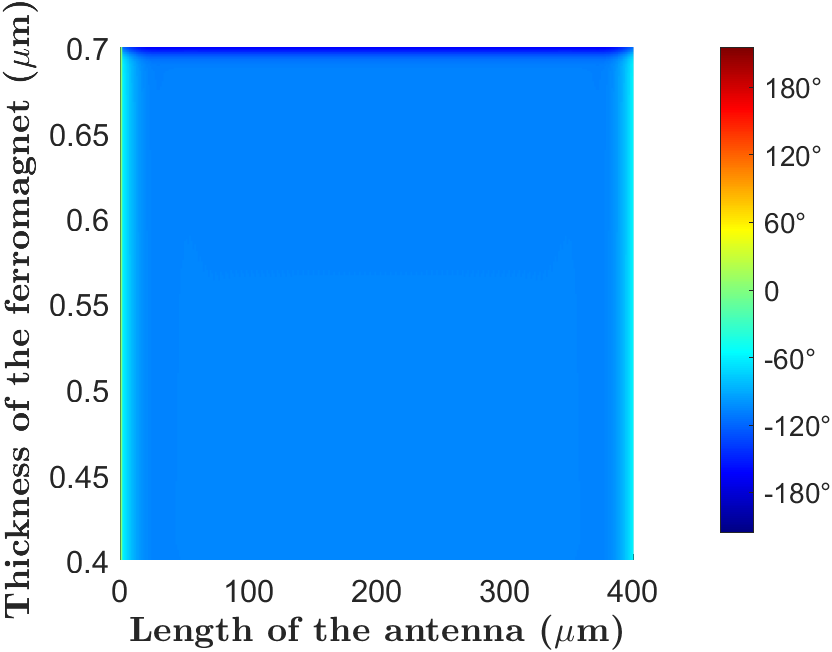}
\end{subfigure}\hspace{5pt}
\begin{subfigure}{0.41\linewidth}
  \centering
  \caption{}
  \label{fig:sfig4}
  \includegraphics[width={\linewidth}]{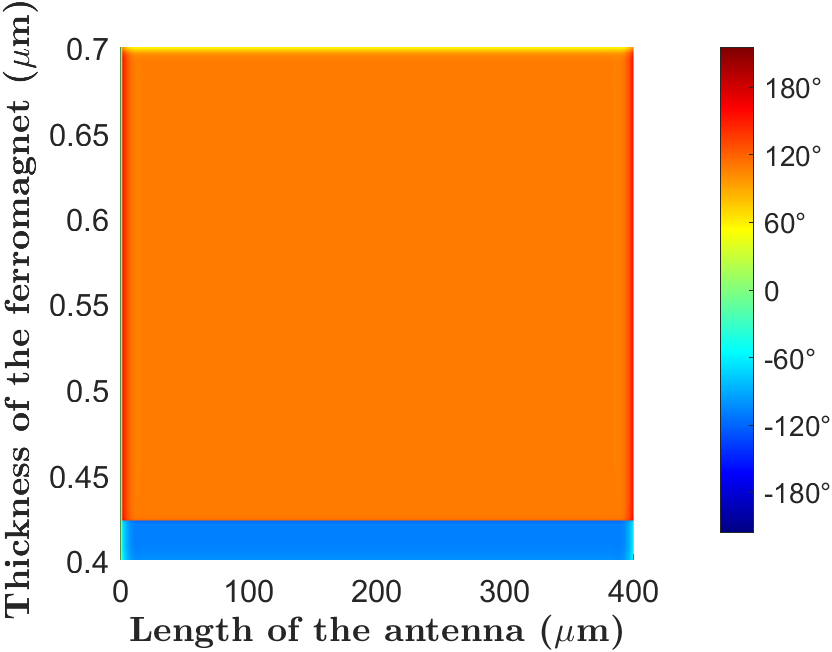}
\end{subfigure}
\caption{Magnitude (a),(b) and phase (c),(d) of the time-harmonic standing spin wave $\varphi_h$ at the first and second resonances.}
\label{fig:fig}
\end{figure*}

\subsection{Standing spin waves}

The LL equation describes the magnetization precession around an effective magnetic field $\mathbf{H_{eff}}$, which can be decomposed into several components:
\begin{equation}
    \mathbf{H_{eff}}=\mathbf{H_0}+\mathbf{H_{exc}}+\mathbf{H_{ani}}+\mathbf{H_{me}}
\end{equation}
where $\mathbf{H_0}=H_0\Vec{x}$ is the bias magnetic field oriented along the x-axis in this study. $\mathbf{H_{exc}}$ the exchange interaction field, $\mathbf{H_{ani}}$ the magnetocrystalline anisotropy field, and $\mathbf{H_{me}}$ the magnetoelastic field.

For thin ferromagnetic films, the demagnetizing field along the thickness direction forces the magnetization to lie in the xy-plane. In the absence of a strong out-of-plane external magnetic field that would cancel the demagnetizing effect, the magnetization unit vector $\mathbf{m}$
has a weak z component. We will neglect this component and consider the magnetization unit vector to lie entirely in the xy-plane:
\begin{equation}
    \begin{split}
        \mathbf{m}&=m_x \vec{x}+m_y\Vec{y}\\
        &=\cos\varphi\vec{x}+\sin\varphi\vec{y}
    \end{split}
\end{equation}
where $\varphi$ is the angle between $\mathbf{m}$ and the x-axis. For a cubic crystal, the exchange interaction field, the magnetocrystalline anisotropy field, and the magnetoelastic field are written respectively:
\begin{equation}
    \begin{split}
        &\mathbf{H_{exc}}=\frac{2A_{exc}}{\mu_0 M_s}\left(\nabla^2m_x\Vec{x}+\nabla^2m_y\Vec{y}\right)\\
        &\mathbf{H_{ani}}=-\frac{2K_1}{\mu_0 M_s}\left(m_x(1-m_x^2)\Vec{x}+m_y(1-m_y^2)\Vec{y}\right)\\
        &\mathbf{H_{me}}=\frac{6\lambda_{111}}{\mu_0 M_s}\sigma_{xz}m_x\Vec{z}
    \end{split}
\end{equation}
$A_{exc}$ is the exchange stiffness constant, $K_1$ the cubic anisotropy constant and $\lambda_{111}$ the magnetostriction constant along the crystallographic direction $[111]$. The projection on the x or y-axis of the LL equation gives the partial differential equation satisfied by the angle $\varphi$:
\begin{eqnarray}
    \frac{\partial \varphi}{\partial t}&&=\frac{2\alpha_L \gamma A_{exc}}{ M_s}\nabla^2\varphi-\frac{\alpha_L\gamma K_1}{2M_s}\sin4\varphi-\alpha_L\mu_0\gamma H_0\sin\varphi\nonumber\\
    &&+\frac{6\gamma\lambda_{111}}{M_s}\sigma_{xz}\cos\varphi
\end{eqnarray}

Acoustically actuated magnetoelectric antennas radiate electromagnetic waves through a perturbation of the magnetization \cite{Luong}. The magnetization is biased along the x-axis, and the dominant static magnetization is along the x-axis. The magnetization unit vector can therefore be linearized as:
\begin{equation}
    \mathbf{m}=\Vec{x}+\varphi\Vec{y}
\end{equation}

The linearization of the magnetization vector also linearizes the LL equation:
\begin{eqnarray}
    \frac{\partial \varphi}{\partial t}&&=\frac{2\alpha_L\gamma A_{exc}}{M_s}\nabla^2\varphi-\alpha_L\gamma\left(\frac{2 K_1}{ M_s}+\mu_0 H_0\right)\varphi\nonumber\\
    &&+\frac{6\gamma \lambda_{111}}{ M_s}\sigma_{zx}(x,z)
\end{eqnarray}

The boundary condition verified by $\varphi$ at $z=z_1$ and $z=z_2$ is:
\begin{equation}
    \nabla_n\cdot \mathbf{m}=\frac{\partial\varphi}{\partial z}=0
\end{equation}
where $n=z$ is the vector normal to the xy-plane. Moreover, the ferromagnetic layer thickness is much smaller than the lateral direction $t_f\ll L$, so the boundary condition at the lateral boundaries $x=0$ and $x=L$ can be approximated by a magnetic wall \cite{Demidov}:
\begin{equation}
    \varphi=0
\end{equation}

The solution of (33) can be decomposed into a time-harmonic term that corresponds to the magnetoelastic contribution and a decay term. The time-harmonic term only plays a role in the radiation and is given by (See Supplemental Material \cite{Supp}):
\begin{widetext}
    \begin{equation}
\begin{split}
    \varphi_{h}&=\frac{6\gamma\lambda_{111}}{M_s}\sum_{n=1}^\infty\sigma_n \Bigg\{\frac{\cos\left(k_ft_f\right)-1}{k_ft_f}\frac{\sin\left(k_n x\right)e^{j\omega t}}{j\omega+\alpha_L\gamma\left(\frac{2 K_1}{ M_s}+\mu_0 H_0+\frac{2 A_{exc}}{M_s}k_n^2\right)}\\
    &+\frac{2}{k_ft_f}\sum_{k=1}^\infty\frac{\left(-1\right)^k\cos\left(k_ft_f\right)-1}{1-\left(\frac{k\pi}{k_ft_f}\right)^2}\frac{\sin\left(k_n x\right)\cos\left(\frac{k\pi(z-z_2)}{t_f}\right)e^{j\omega t}}{j\omega+\alpha_L\gamma\left(\frac{2 K_1}{ M_s}+\mu_0 H_0+\frac{2 A_{exc}}{M_s}\left(k_n^2+\left(\frac{k\pi}{t_f}\right)^2\right)\right)}\Bigg\}
\end{split} 
\end{equation}
\end{widetext}
where $\sigma_n$ is the magnitude of the stress in the ferromagnetic layer in (27). The spin wave is complex-valued, implying that both magnitude and phase are used to describe it. 
\begin{table}[b]
\caption{\label{tab:table2}
Material properties in the multiferroic antenna.}
\begin{ruledtabular}
\begin{tabular}{ll}
 \multicolumn{2}{c}{Ni \cite{Handley}}\\
\hline
$M_s$ (A/m) & $4.8\times 10^5$  \\
$\mu_0 H_0$ (mT) & 30 \\
$A_{exc}$ (J/m) & $1.05\times 10^{-11}$  \\
$\lambda_{111}$ & $-24\times 10^{-6}$ \\
$K_1$ ($\mathrm{J/m^3}$)& $-4.5\times 10^3$ \\
$\left(C_{11}^f,C_{12}^f,C_{44}^f \right)$ (GPa)& $\left(250,160,118\right)$   \\
$\rho$ ($\mathrm{kg/m^3}$)& 8900  \\
$\epsilon_r$ & 1 \\
$\alpha_L$ & 0.025 \\
$\eta_s$ & 0.001 \\\hline
\multicolumn{2}{c}{AlN \cite{Comsol}} \\
\hline
$e_{15}$ ($\mathrm{C/m^2}$)& -0.48  \\
$\left(C_{11}^p,C_{13}^p,C_{33}^p,C_{44}^p \right)$ (GPa)& $\left(410,99,389,125\right)$   \\
$\rho$ ($\mathrm{kg/m^3}$)& 3300  \\
$\epsilon_r$ & 9 \\
$\tan\delta$&0.01\\
$\eta_s$ & 0.001 \\
\hline
\multicolumn{2}{c}{Pt \cite{Comsol}} \\
\hline
$\sigma$ (S/m)& $8.9\times 10^6$ \\
$\rho$ ($\mathrm{kg/m^3}$)& 21450 \\
E  (GPa)&168 \\
$\nu$& 0.38\\
$\eta_s$ & 0.001 \\
\end{tabular}
\end{ruledtabular}
\end{table}

Fig. 3 shows the spatial distribution of the magnitude and phase of the time-harmonic standing spin wave in the ferromagnetic layer at the first and second mechanical resonances. As expected from (33), the spatial distribution is the same as the shear stress. At the first resonance, the magnitude decreases along the thickness of the ferromagnetic layer from $2\times 10^{-3}$ at $z=400~\mathrm{nm}$ to 0 at the top of the ferromagnetic layer $z=700~\mathrm{nm}$. At the second resonance, the magnitude is constant in the middle of the ferromagnet with a normalized value of $1.2\times 10^{-3}$. Figs. 3(c)-3(d) show that the spin wave at the second resonance has an opposite phase compared to the first resonance in the upper part of the ferromagnetic layer. It corresponds to the change of polarity after each zero of the displacement vector for a thickness-shear mode.

The spin waves being calculated here for both space and time dependencies, enables as shown further, to obtain an analytical expression of the electric vector potential in the far-field without the use of the infinitesimal dipole approximation. Moreover, since the electric vector potential is a volume integral of the time-derivative of the magnetization, by examining the stress and spin wave spatial distribution, we expect the second mode to radiate more than the first mode.

\subsection{Electromagnetic waves radiation }

Recalling the definition of the magnetic field $\mathbf{B}=\mu_0\left(\mathbf{H}+\mathbf{M}\right)$, in a ferromagnetic material, a time variation of the magnetization, equivalent to a time variation of the magnetic field, generates an electric field through the Maxwell-Faraday equation. The radiation source is:
\begin{equation}
        \mu_0\frac{\partial \mathbf{M}}{\partial t}\approx \mu_0 M_s\frac{\partial\varphi }{\partial t}\Vec{y}
\end{equation}
where we used the small angle approximation (32). The radiation source is a magnetic current source along y. One obtains the EM field through the computation of the associated electric vector potential in the volume of the ferromagnet. In the far field zone: 
\begin{equation}
    \mathbf{F_e}=\frac{\epsilon_0\mu_0 M_s}{4\pi r}e^{-j\mathrm{k_0}r}\Vec{y}\int_0^L\int_0^{W_f}\int_{z_1}^{z_2}\frac{\partial\varphi }{\partial t}e^{j\mathrm{k_0}r^\prime\cos\psi}dv^\prime
\end{equation}
where $W_f$ is the width of the ferromagnetic layer, $r$ the distance to the observation point, $r^\prime$ spans the ferromagnet volume and $\mathrm{k_0}$ is the free-space wave number. $\psi=\frac{\vec{r}\cdot\Vec{r^\prime}}{r}$ is the angle between the vector that goes from the origin of the coordinate system to the observation point and the vector that goes from the origin to the source point. (See Supplemental Material \cite{Supp} for the explicit expression of the vector potential). Therefore, it is possible to obtain a closed-form solution of the radiated EM field whose components are expressed by:
\begin{equation}
    \begin{split}
        &E_\theta=-j\omega Z_0 \mathbf{F}_e\cdot \vec{e_\phi}\\
        &E_\phi=j\omega Z_0\mathbf{F}_e\cdot \vec{e_\theta}
    \end{split}
\end{equation} 
where $Z_0=\sqrt{\mu_0/\epsilon_0}$ is the vacuum impedance and $\vec{e_\phi},\vec{e_\theta}$ the unit vectors in spherical coordinates. The radiation power is given by:
\begin{equation}
    P_{rad}=\int_0^{2\pi}\int_0^\pi\frac{\left|E_\theta\right|^2+\left|E_\phi\right|^2}{2Z_0}r^2\sin\theta d\theta d\phi
\end{equation}

The radiation efficiency is calculated by:
\begin{equation}
    \eta_{rad}=\frac{P_{rad}}{P_{in}}\times 100\%
\end{equation}
where $P_{in}$ is the accepted power by the multiferroic antenna, expressed in terms of the input admittance $Y_{in}$, the applied voltage $V$, and the system impedance $Z_s$ as \cite{Yao}:
\begin{equation}
    P_{in}=\frac{V^2}{2}\left\lvert Y_{in}^*-\frac{1}{Z_s}\right\rvert
\end{equation}

Fig. 4(a) shows the radiated EM field of the multiferroic antenna in the xz-plane and yz-plane for the first two resonances. The planar dimensions of the multiferroic antenna are $400\mu m\times 400\mu m$. The radiated field intensity at the second resonance is approximately two times that of the first resonance. Fig. 4(b) shows the radiated power and radiation efficiency at the two resonances. The radiated power of the first and second resonances are $4.57\times 10^{-14}$ W and 0.16 pW respectively, and the corresponding radiation efficiencies are $5\times 10^{-10}\%$ and $1.7\times 10^{-9}\%$. Indeed, the maximum values of the stress in the ferromagnetic layer of 0.35 MPa and 0.42 MPa for the first and second resonances, respectively, already indicated a higher radiation from the second resonance. However, as will be discussed in the subsequent part, the thicknesses of the ferromagnetic and piezoelectric layers were not optimized for maximum radiation power.

\begin{figure}[htp!]
\begin{subfigure}{\linewidth}
  \centering
  \caption{}
  \label{fig:sfig1}
  \includegraphics[width=\linewidth]{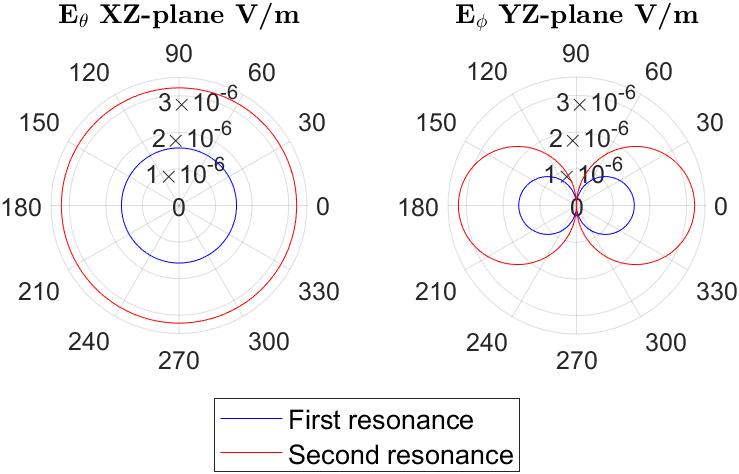}
\end{subfigure}
\hfill
\begin{subfigure}{\linewidth}
  \centering
  \caption{}
  \label{fig:sfig3}
  \includegraphics[width=\linewidth]{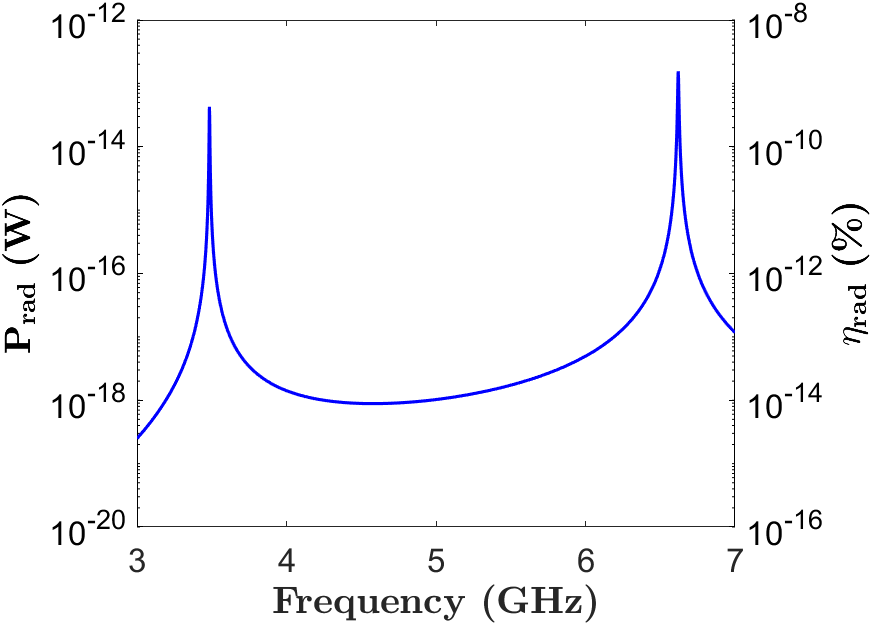}
\end{subfigure}
\caption{(a) Simulated far-field radiation patterns at the first and second resonances. (b) Radiation power (left axis) and radiation efficiency (right axis) at the first and second resonances.}
\label{fig:fig}
\end{figure}

\section{Optimization and Akhiezer limit}
\begin{table}[b]
\caption{\label{tab:table2}
Average Gr{\"u}neisen's parameter and $f\times Q_a$ product for the different piezoelectric materials.}
\begin{ruledtabular}
\begin{tabular}{ccc}
 &$\gamma_{avg}$ \cite{Ghaffari}&$f\times Q_a$\\
\hline
AlN & 0.91 & $2.5\times 10^{13}$ \\
$\mathrm{LiTaO_3}$& 2.37 & $2.9\times 10^{13}$\\
$\mathrm{LiNbO_3}$& 2.37 & $3.086\times 10^{13}$\\
\end{tabular}
\end{ruledtabular}
\end{table}

\begin{figure*}[htp!]
\begin{subfigure}{0.42\linewidth}
  \centering
  \caption{}
  \label{fig:sfig1}
  \includegraphics[width=\linewidth]{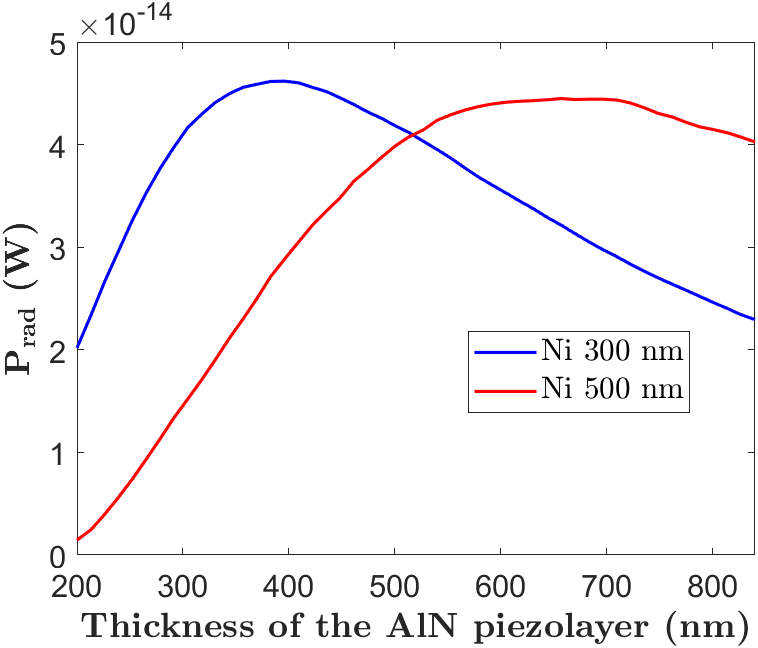}
\end{subfigure}\hspace{5pt}
\begin{subfigure}{0.43\linewidth}
  \centering
  \caption{}
  \label{fig:sfig3}
  \includegraphics[width=\linewidth]{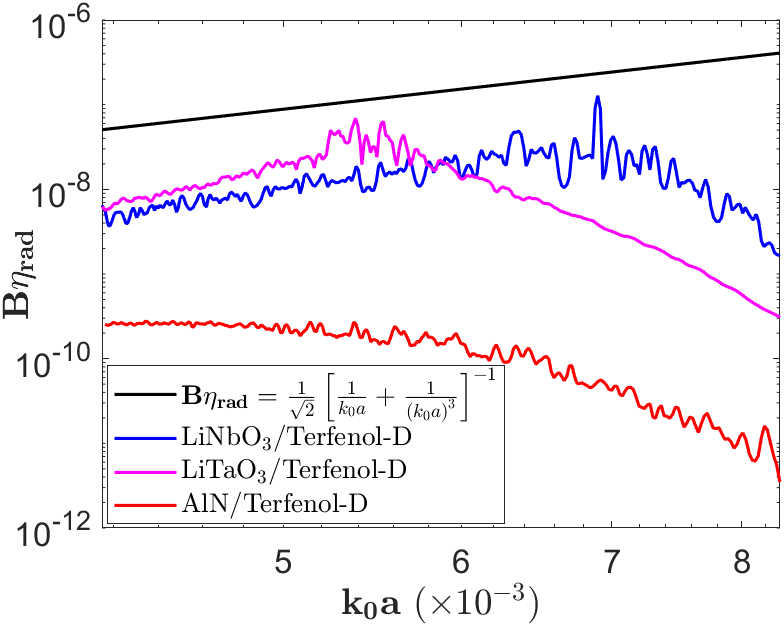}
\end{subfigure}
\caption{(a) Radiated power as a function of the thickness of the AlN piezolayer for two thicknesses of the Ni layer. (b) Comparison with Chu's limit of the efficiency-bandwidth product of different multiferroic antennas when the Akhiezer damping is considered. $\mathrm{k_0a}$ is the product of the free-space wavenumber $\mathrm{k_0}$ and the minimum radius a, of a sphere enclosing the antenna.}
\label{fig:fig}
\end{figure*}

We show in this section the impact of the viscosity constant and the thicknesses of the piezolayer and ferromagnetic layer, on the performance of the multiferroic antenna.  Fig. 5(a) plots the radiated power as a function of the thickness of the AlN piezolayer for a thickness of the Ni layer of 300 and 500 nm. For a thickness of the Ni layer of 300 nm, the maximum radiated power is obtained for a thickness of the AlN piezolayer of 380 nm. Moreover, a comparison of the maximum radiated power by the 300 and 500 nm cases, reveals that increasing the thickness of the Ni layer decreases the value of the maximum radiated power from $4.7\times 10^{-14}$W to $4.5\times 10^{-14}$W, respectively. 

Fig. 5(b) compares the efficiency-bandwidth product of different multiferroic antennas with Chu's limit. Terfenol-D is used as ferromagnetic material for its giant magnetostriction constant $\lambda_{111}=1.6\times 10^{-3}$ and different piezoelectric materials including AlN, $\mathrm{LiNbO_3}$ and $\mathrm{LiTaO_3}$ are compared. The viscosity constant considered is the minimum achievable viscosity constant due to internal frictions, denoted as $\eta_{min}$. Internal frictions are related in the GHz regime to the phonon-phonon scattering, also referred to as Akhiezer damping, which limits the maximum achievable mechanical quality factor $Q_a$ for a single crystal-resonator \cite{Akhiezer,Ghaffari,Candler}:
\begin{equation}
    Q_a=\frac{\rho v_{a}^4}{\gamma_{avg}^2\kappa T \omega}
\end{equation}
where $\rho$ is the mass density, $v_a$ the average Debye acoustic velocity, $\gamma_{avg}$ the average Gr{\"u}neisen's parameter, $\kappa$ the thermal conductivity, $T$ the ambient temperature and $\omega$ the frequency of the AC voltage applied to the electrodes. For simplicity, we consider the $Q_a$ of the bilayer structure to be the $Q_a$ of the piezolayer. The minimum achievable viscosity constant is calculated taking the inverse of the quality factor $\eta_{min}=Q_a^{-1}$. To vary the resonance frequency of the multiferroic antennas, we vary the thickness of the piezolayers from 100 to 1000 nm, with a fixed thickness of the Terfenol-D layer of 500 nm.

At low frequencies, the viscosity constant is low and the bandwidth-efficiency product increases until reaching the maximum value allowed by the structure. Once the maximum value is reached, the bandwidth-efficiency product decreases along with the increase of the resonance frequency, and thus along with the increase of the viscosity constant.

The AlN case has a very low efficiency-bandwidth product because of its low experimental effective electromechanical coupling coefficient $k_t^2$:
\begin{equation}
    k_t^2=\frac{\pi}{2}\frac{f_s}{f_p}\cot\left(\frac{\pi}{2}\frac{f_s}{f_p}\right)
\end{equation}
where $f_s$ and $f_p$ are the series and parallel resonances. $\mathrm{LiNbO_3}$ and $\mathrm{LiTaO_3}$ with higher $k_t^2$ show efficiency-bandwidth products close to Chu's limit, with values of $1.27\times 10^{-7}$ and $6.9\times 10^{-8}$ at $\mathrm{k_0a}=6.9\times 10^{-3}$ and $\mathrm{k_0a}=5.38\times 10^{-3}$, respectively. As shown in Fig. 5(b), the choice of $\mathrm{LiNbO_3}$ or $\mathrm{LiTaO_3}$ enhances the radiation efficiency by approximately two orders of magnitudes compared to AlN. 

\section{Conclusions}

This paper demonstrates a two-dimensional closed-form analytic model of an XBAR multiferroic antenna. An extended Mason model to two dimensions is proposed to account for the particle displacement along the thickness and the lateral directions. The extended Mason model enables not only to obtain the electrical response of the structure, as the conventional Mason model would do in one dimension, but also allows to obtain the mechanical response through the mode shape. This theory, developed here based on a bilayer piezoelectric-ferromagnet heterostructure, is also applicable to structures with more than two layers. Indeed, adding extra layers of buffer non-magnetic materials only modifies the equivalent impedances seen by the piezoelayer at its top and bottom surfaces, shifting the resonant frequencies. The displacement vector in the ferromagnetic layer, needed for the computation of the electric vector potential, can be obtained similarly as the two-layer case, by application of continuity conditions on the displacement vectors.

Moreover, using a unidirectional coupling between the governing equations, we found closed-form solutions for the spin wave and radiated EM field without the use of infinitesimal dipole approximation. The dependencies between the antenna response and the parameters that compose it are explicitly specified. As a consequence, optimization of these parameters can readily be conducted for the study of radiation enhancement through the definition of the adequate thickness of the piezoelectric and ferromagnet layers leading to the maximum radiated power. 

Finally, the efficiency-bandwidth product of different multiferroic antennas is compared with Chu's limit. The comparison is conducted with the highest achievable mechanical quality factor in the piezolayer, known as the Akhiezer limit. We show that at this fundamental mechanical limit, the efficiency-bandwidth products of $\mathrm{LiNbO_3}$/Terfenol-D and $\mathrm{LiTaO_3}$/Terfenol-D multiferroic antennas approach Chu's limit. Furthermore, the choice of Terfenol-D ferromagnetic material combined with $\mathrm{LiNbO_3}$ or $\mathrm{LiTaO_3}$ piezoelectric materials allows a radiation efficiency enhancement of approximately two orders of magnitudes compared to AlN/Terfenol-D. Thus, the fundamental factor limiting radiation efficiency is Chu's limit. However, when considering the coupled physics nature of multiferroic antennas, phonon-phonon scattering in the Akhiezer regime is most likely to limit the antenna radiation for GHz devices, and not Chu's limit.

\section*{Acknowledgments}
This work was supported by the National Science and Technology Council, Taiwan, under contracts: NSTC 112-2918-I-002-011, 112-2218-E-002-031, and 112-2119-M-002-015.
% The \nocite command causes all entries in a bibliography to be printed out
% whether or not they are actually referenced in the text. This is appropriate
% for the sample file to show the different styles of references, but authors
% most likely will not want to use it.
\nocite{*}

\bibliography{Reference}% Produces the bibliography via BibTeX.

\end{document}